\documentclass[letter,amsmath,amssymb,floatfix]{revtex4}

\usepackage{graphicx}% Include figure files
\usepackage{dcolumn}% Align table columns on decimal point
\usepackage{bm}% bold math
\usepackage{amssymb}
\usepackage[lflt]{floatflt}

\begin{document}

\title{Thermal Dimuon Yields at NA60}
\author{K. Dusling}
\author{D. Teaney}
\author{I. Zahed}
\affiliation{Department of Physics \& Astronomy, State University 
of New York, Stony Brook, NY 11794-3800, U.S.A.}

\date{\today}  

\begin{abstract}  
Dilepton emission rates from a hadronic gas at finite temperature and
baryon density are completely constrained by broken chiral symmetry in
a density expansion. The rates can be expressed in terms of vacuum 
correlations which are measured in $e^+e^-$ annihilation, $\tau$ decays
and photo-reactions on nucleons and nuclei~\cite{paper1, paper2, paper3}.  
In this paper, the theoretical results are summarized and the total 
dimuon yield is calculated by integrating the dimuon rates over the 
space-time history of a fireball based on hydrodynamic calculations with
CERN SPS conditions. The resulting dimuon yield is in good agreement
with the recent measurements reported by NA60~\cite{sanja05}. 

\end{abstract}


\maketitle  

\section{Introduction}

It is expected that above a critical temperature, $T_c\approx 170$ MeV, QCD 
undergoes a chiral phase transition where the relevant degrees of freedom 
change from mesons and baryons to a phase of strongly coupled quarks and 
gluons, the strongly coupled quark-gluon plasma (sQGP). This new phase of 
matter is being searched for in a number of ultra-relativistic heavy ion 
facilities.  There are a number of current observations in favor of the sQGP, 
ranging from hydrodynamical flow (soft probes) to jet quenching (hard probes). 
However, most of these observations are blurred by the fact that the
space-time evolution of the sQGP is short and its conversion to hadronic
matter is involved. Since the latter dominates the final stage of the
evolution, it is producing competing signals that interfere with those from
the sQGP. In this respect, dilepton and photon emissions are interesting
probes of the collision region as neither interact strongly with the medium 
produced in these collisions, thus they probe the {\em early} stages of the 
collision.  This is in contrast to hadronic observables which thermalize 
along with the collision region thus providing information only on the 
late (or freeze-out) stage of the collision. 

Making quantitative predictions of the production rates of dileptons and 
photons is difficult for a number or reasons.  Since the temperature produced 
in typical heavy-ion collisions is in the range of 200-300 MeV which is 
about the QCD scale factor, $\Lambda$, the differential cross sections can not
be computed in a weak-coupling expansion.  Another uncertainty is detailed 
knowledge of the evolution of both hadronic matter and quark gluon phase 
produced in heavy-ion collisions.  In addition there is also a background of 
dileptons from other processes not occurring in the quark-gluon plasma such 
as hadronic decays. 

In the past there have been a number of experiments probing photons and 
dileptons created in hadronic collisions.  One of the most recent experiments 
was the CERES (NA45) taking place as the CERN SPS collider which looked for 
dielectrons.  It was found that the dielectron production exceeded the 
theoretical expectations for {\em conventional} processes in both hadronic 
and QGP matter~\cite{HungShuryak96}, especially in the mass region 
$0.3\leq M \text{(GeV)} \leq 0.6$~\cite{CERES95}. A number of theoretical
analyses were put forward to explain this excess based on effective
Lagrangians with medium modification~\cite{Rapp,LiGale98} and dropping
vector meson masses~\cite{BrownRho}. Model independent emission rates
constrained by the strictures of broken chiral symmetry and data were
unable to account for the excess rate reported by NA45
\cite{paper1,paper2,paper3}. However, the large statistical and systematic 
errors reported by NA45 in exactly the excess region, did not allow for a 
definitive conclusion as to the theoretical nature of the emissivities.

In this letter we revisit these issues in light of the recently reported
dimuon data from the NA60 collaboration using In-In collisions at 
158 Gev/Nuc~\cite{sanja05}. These data have far better statistics, which gives much better
constraints on any medium modification to the vector mesons \cite{Hees06, Renk06}.  We use the model independent analysis
in~\cite{paper1,paper2,paper3} to analyze these data, whereby the
emissivities are constrained by broken chiral symmetry in a dilute
hadronic medium, and by non-perturbative QCD in the sQGP. The collision
expansion and composition are extracted from an underlying hydrodynamical 
evolution set to reproduce the CERN SPS conditions.

\section{Dilepton Emission Rates from a Fireball}

The rate of dilepton emission per unit four volume for particles in 
thermal equilibrium at a temperature T is related to the thermal 
expectation value of the electromagnetic current-current 
correlation function~\cite{McLerran, Weldon}.  For massless leptons 
with momenta $p_1$ and $p_2$, the rate per unit invariant momentum 
$q=p_1+p_2$ is given by:

\begin{equation}
\label{eq:rate}
\frac{dR}{d^4q}=\frac{-\alpha^2}{3\pi^3 q^2}\frac{1}
{1+e^{ q^0/T}}\text{Im}{\bf W}^F(q)
\end{equation}

where $\alpha=e^2/4\pi$, T is the temperature and 

\begin{equation}
\label{eq:WF}
{\bf W}^{F}(q)= i\int{d^4x}\text{ } 
e^{iq\cdot x}\text{Tr}\biglb[e^{-({\bf H}-\mu_B{\bf N}-\Omega)/T} 
T^* {\bf J}^\mu(x) {\bf J}_\mu(0) \bigrb]
\end{equation}
where $e{\bf J}_\mu$ is the hadronic part of the electromagnetic current, 
{\bf H} is the hadronic Hamiltonian, $\mu_B$ is the baryon chemical 
potential, {\bf N} is the baryon number operator, and $\Omega$ is the 
Gibbs energy.  The trace is over a complete set of hadron states.

In order to take into account leptons with mass $m_l$ the right-hand side 
of Eq.~\ref{eq:rate} is multiplied by

\begin{equation}
\label{eq:lep_mass}
(1+\frac{2m^2_l}{q^2})(1-\frac{4m^2_l}{q^2})^{1/2}
\end{equation}
To compare the theoretical dilepton production rates with those 
observed in heavy ion collisions, the rates must be integrated over 
the space-time history of the collision region and then finally 
integrated over the dilepton pair's transverse momentum and rapidity 
in order to compare with the yields measured by the NA60 collaboration.  
The final expression for the rates is given as:

\begin{equation}
\label{eq:acc}
\frac{dN}{dM}=2\pi M\int{dy}\int{dq_\perp}\cdot 
q_\perp\times Acc(M,q_\perp,y)\int_{\tau_0}^{\tau_{f.o}}
{\tau d\tau}\int_{-\infty}^{\infty}{d\eta}\int_{0}^{r_{max}}{rdr}\int_{0}^{2\pi}{d\theta} \frac{d^4R}{d^4q d^4x}\biglb(M,|\vec{q}|, 
T,\mu_B,x\bigrb)
\end{equation}

where $M=\sqrt{q^2}$ is the dilepton invariant mass, y is the dilepton 
pair rapidity, $\eta$ is the spatial rapidity, $q_\perp$ is the dilepton pair transverse momentum (with $\theta$ defined as the angle between $q_\perp$ and the fluid element's velocity), x is 
the hadron fraction, r is radial coordinate (with $r_{max}$ set by the freeze-out temperature), and
$Acc(M,q_\perp,y)$ is the experimental acceptance taking into account 
that the CERES detector covers a limited rapidity in the interval 
$y=2.9-4.5$ in the lab frame.  

The integration over $\eta, r, \theta$ and $\tau$ was done over the full hydrodynamic simulation of the fireball as described below.  $|\vec{q}|$ can be found by considering the two invariants $q^\mu q_\mu=M^2$ and $u^\mu q_\mu$ constructed from the dilepton momentum and fluid 4-velocity which can be expressed as:

\begin{align}
\label{eq:mtm_4vel}
&q^\mu = \biglb(M_\perp \cosh(y),q_\perp,M_\perp \sinh(y) \bigrb) \nonumber\\
&u^\mu = \biglb(\gamma_\perp \cosh(\eta), \gamma_\perp v_\perp, \gamma_\perp \sinh(\eta) \bigrb)
\end{align}

giving

\begin{equation}
\label{eq:qmag}
|\vec{q}| = \biglb[-M^2+\biglb(\gamma_\perp M_\perp \cosh(\eta)-u_\perp q_\perp\cos(\theta)\bigrb)^2\bigrb]^{1/2}
\end{equation}

where $u_\perp=\gamma_\perp v_\perp$, $\gamma_\perp=\frac{1}{\sqrt{1-v_\perp^2}}$, and $v_\perp$ is the transverse fluid velocity which is taken from the hydrodynamic simulation.

The acceptance function has a complicated dependence on $M, q_\perp$ and y, but since our rates are y-independent we have used an acceptance with $M$ and $q_\perp$ dependence built to specifications provided by the NA60 collaboration \cite{PC}.  Without detailed hadronic data available (such as transverse mass spectra and HBT analysis) a careful consideration of hadronic input, such as freeze-out temperature, cannot be made.  Therefore there is a large uncertainty in the overall normalization of the yields, which depends strongly on $T_{f.o.}$.  In addition, the assumption of boost invariance can also affect the normalization as the acceptance probes very specific rapidities.  The approach taken here is to normalize our results to the excess data in the peripheral centrality windows which fixes the normalization in the central bins.

\section{Spectrum above T$_C$}

At temperatures T$>$T$_C$ lattice calculations have predicted that 
the relevant degrees of freedom consists of (strongly) interacting
quarks and gluons.  In order to compute the dilepton production rates as 
one would expect from a conventional phase of quark-gluon plasma we use 
the Born q\={q} annihilation term \cite{cleymans, bellac} which for massless 
quarks is

\begin{equation}
\label{eq:Bornqqbar}
\text{Im} {\bf W}^R=\frac{1}{4\pi}\Biglb( N_C 
\sum_{q=u,d,s}e^2_q \Bigrb) q^2 \Biglb[ 
1+\frac{2T}{|\Vec{q}|}\ln(\frac{n_+}{n_-}) \Bigrb]
\end{equation}
where $N_C$ is the number of colors, $e_q$ the charge of the quarks, 
and $n_\pm=1/(e^{(q_0\pm|\Vec{q}|)/2T}+1)$.  It should be mentioned 
that~\ref{eq:Bornqqbar} reduces to the well-known vacuum result 
$(\text{Im} W^R=\frac{-q^2 N_C}{4\pi}\sum_q e_q^2)$ at T=0 and that the 
finite temperature rate is always smaller then the T=0 rate due to 
Pauli blocking.  

\begin{figure}
\includegraphics[scale=.75]{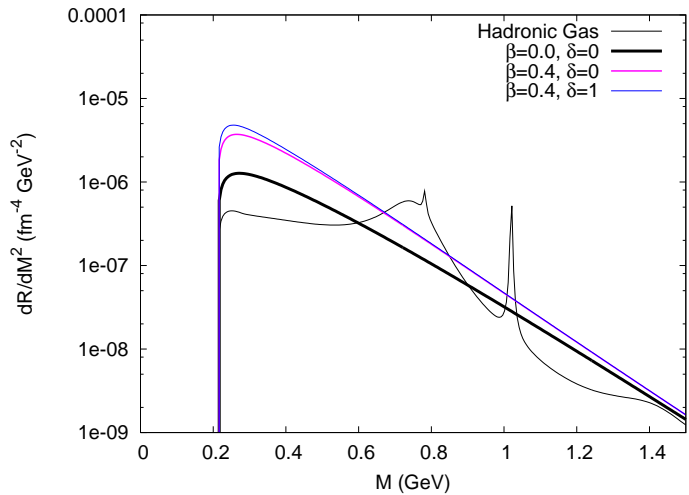}
\caption{Integrated dimuon rates from the plasma phase for T=150 MeV.  
The thick solid line shows the perturbative q\={q} annihilation rates 
while the this solid lines show the results for 
non-vanishing A$^2_4$, B$^2$ and E$^2$ condensates and for 
only a non-vanishing A$^2_4$ condensate.  
For comparison the integrated 
hadronic rate at T=150 MeV and $\mu_B=225$ MeV is also shown, which will 
be discussed in the next section.}
\label{fig:dimuon_rates_qgp}
\end{figure}

It has also been seen in lattice simulations that near the critical 
temperature $T_c$ there are still substantial chromoelectric and 
chromomagnetic condensates present leading to additional 
non-perturbative effects.  It was shown in \cite{LWZH98} that the 
enhancement to the dilepton rates in a plasma with non-vanishing 
chromoelectric and chromomagnetic condensates can be given by

\begin{equation}
\label{eq:nonpert}
\text{Im} {\bf W}^R=\frac{1}{4\pi}\Biglb( N_C 
\sum_{q=u,d,s}e^2_q \Bigrb) \Biglb[ q^2 
\Biglb< \frac{\alpha_s}{\pi}A^2_4 \Bigrb>-
\frac{1}{6}\Biglb< \frac{\alpha_s}{\pi}E^2\bigrb>+
\frac{1}{3}\Biglb< \frac{\alpha_s}{\pi}B^2\Bigrb>  \Bigrb] 
\biglb( \frac{4\pi^2}{T|\Vec{q}|}\Bigrb) 
\Biglb( n_+(1-n_+)-n_-(1-n_-) \Bigrb)
\end{equation}
where the values of the above condensates in~\ref{eq:nonpert} can only 
be estimated using non-perturbative calculations such as lattice QCD.  
The net result is a substantial enhancement (as seen in 
Fig.~\ref{fig:dimuon_rates_qgp} for the case of dimuons) of the dilepton 
production rates below an invariant mass of $\approx 0.4$GeV.  For the 
remainder of the paper we refer to the perturbative plasma rates as those 
given by~\ref{eq:Bornqqbar} and the non-perturbative plasma rates as those 
given the sum of equations~\ref{eq:Bornqqbar} and~\ref{eq:nonpert} using 
$\delta\equiv \biglb<\frac{\alpha_s}
{\pi}E^2\bigrb>/(200 MeV)^4=\biglb<\frac{\alpha_s}
{\pi}B^2\bigrb>/(200 MeV)^4=0$ and 
$\beta\equiv \biglb<\frac{\alpha_s}{\pi}A^2_4\bigrb>/T^2=0.4$ 
which is the lower of the non-perturbative curves in 
Fig.~\ref{fig:dimuon_rates_qgp}.  An explanation of the choice 
of $\beta=0.4$ can be found in \cite{LWZH98}.

%pl_had_muon_muB225_T150_drdM2
%drdM2plot.gp

\section{Spectrum below T$_C$}

Even though there are various approaches to calculating production 
rates, they differ in the way in which the current-current correlation 
function in Eq.~\ref{eq:rate} is approximated and evaluated.  
The approach taken here is to use a chiral reduction formalism in 
order to reduce the current-current correlation function in~\ref{eq:WF} 
into a number of vacuum correlation functions which can be constrained 
to experimental $e^+e^-$ annihilation, $\tau$-decay, two-photon fusion 
reaction, and pion radiative decay experimental data.

For temperatures, T $\leq m_\pi$ and for nucleon densities, 
$\rho_N \leq 3\rho_0 $ the trace in Eq.~(\ref{eq:WF}) 
can be expanded in pion and nucleon states.  
Keeping terms up to first order in meson and nucleon 
density gives \cite{paper3}

\begin{equation}
\label{eq:exp}
\text{Im} {\bf W}^F(q)=-3q^2
\text{Im} {\bf \Pi}_V(q^2)+\frac{1}{f^2_a}
\int{da}{\bf W}^F_1(q,k)+\int{dN}{\bf W}^F_N(q,p)
\end{equation}
with phase space factors of

\begin{equation}
\label{eq:phase_space_N}
dN=\frac{d^3p}{(2\pi)^3}\frac{1}{2E_p}\frac{1}{e^{(E_p-\mu_B)/T}+1}
\end{equation}
and

\begin{equation}
\label{eq:phase_space_a}
da=\frac{d^3k}{(2\pi)^3}\frac{1}{2\omega_k^a}\frac{1}{e^{\omega_k^a/T}-1}
\end{equation}

with nucleon and meson energies of E$_p=\sqrt{m^2+p^2}$ and  
$\omega_k^a=\sqrt{m_a^2+k^2}$ respectively.

The first term in~\ref{eq:exp} is the transverse part of the isovector 
correlator $\langle 0|T^*{\bf VV}|0 \rangle$ which can be determined 
experimentally from electroproduction data and gives a result 
analogous to the resonant gas model.  At low and intermediate invariant 
mass the spectrum is dominated by the $\rho(770 MeV)$ and $\rho'(1450 MeV)$.

The term linear in meson density (the second term in Eq.~\ref{eq:exp}) can 
be related to experimentally measured quantities via the three flavor 
chiral reduction formulas \cite{CRF}.  It is shown in~\cite{paper1,paper3} 
that the dominant contribution comes solely from the part involving 
two-point correlators which gives:

\begin{align}
\label{eq:lin_in_meson}
&{\bf W}^F_1(q,k)=\frac{12}{f_\pi^2}q^2\text{Im} 
{\bf \Pi}_V^I(q^2)+\frac{12}{f_K^2}q^2\text{Im} 
\bigglb( {\bf \Pi}_V^I(q^2)+\frac{3}{4}{\bf \Pi}_V^Y(q^2) \biggrb)\nonumber\\
&-\frac{6}{f_\pi^2}(k+q)^2\text{Im} {\bf \Pi}_A^I 
\biglb( (k+q)^2\bigrb)-\frac{6}{f_K^2}(k+q)^2\biglb[\text{Im} 
{\bf \Pi}_A^V\biglb( (k+q)^2\bigrb)+\text{Im} {\bf \Pi}_A^U\biglb( 
(k+q)^2\bigrb) \bigrb] + (q\to -q)\nonumber\\
&+\frac{8}{f_\pi^2}\biglb( (k\cdot q)^2-m_\pi^2 
q^2\bigrb) \text{Im} {\bf \Pi}_V^I(q^2)\times\Re 
\Delta_R^\pi(k+q)+(q\to-q)\nonumber\\
&+\frac{8}{f_K^2}\biglb( (k\cdot q)^2-m_K^2 q^2\bigrb) 
\text{Im} \bigglb({\bf \Pi}_V^I(q^2)+\frac{3}{4}{\bf \Pi}_V^Y\biggrb)
\times\Re \Delta_R^K(k+q)+(q\to -q)
\end{align}
Where $\Delta_R^a$ is the retarded meson propagator given by 
$1/(q^2-m_a^2+i\epsilon)$ and ${\bf \Pi}_A$ is the transverse part of 
the iso-axial correlator $\langle 0|T^*{\bf j}_A{\bf j}_A|0 \rangle$.  The 
spectral functions appearing in Eq.~(\ref{eq:lin_in_meson}) can be 
related to both $e^+e^-$ annihilation as well as $\tau$-decay data 
as was compiled in \cite{Huang95}.  As already shown in \cite{paper1} 
the spectral function can be directly related the form factor, 
${\bf F}_V$, via the KSFR relation where ${\bf F}_V$ is 
parameterized in the common Breit-Wigner form where the 
resonance parameters and decay constants are taken from 
empirical data.  Included in the data are contributions to 
the spectral function from the $\rho, \omega, \phi, a_1, K_1$ 
and some of their radial excitations (see Table I in \cite{paper3}).

\begin{figure}
\includegraphics[scale=.75]{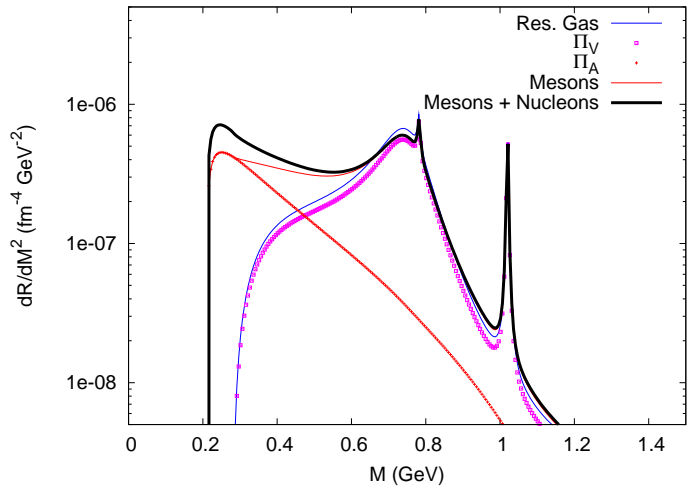}
\caption{The total integrated dimuon rates from a hadronic gas at T=150 MeV.  
The curve labeled Res. Gas shows the analogue of the resonant gas contribution (the first term in 
Eq. \ref{eq:exp}).  The points labeled ${\bf \Pi}_V$ and ${\bf \Pi}_A$ 
give the contributions from all of the respective spectral functions 
in equations \ref{eq:exp} and \ref{eq:lin_in_meson}.  The thin line 
labeled meson is the total rate given by a hadronic gas without 
nucleons.  The thick solid line gives the total dimuon rate when 
nucleons  (shown here for $\mu_B=225$ MeV) are taken into account.} 
\label{fig:dimuon_rates_had}
\end{figure}

It can be seen in Fig.~\ref{fig:dimuon_rates_had} that the term 
linear in meson density decreases the rates from the resonance 
gas contribution for the mass region above the two pion threshold.  
However below the two pion threshold the only contribution to the 
rates come from the ${\bf \Pi}_A$ terms in Eq.~\ref{eq:lin_in_meson}.  
This is because the axial spectral density is integrated over 
all momentum in the thermal averaging (Eq.~\ref{eq:exp}), 
which weakens the $(k+q)^2$ factor in Eq.~\ref{eq:lin_in_meson} 
allowing the $1/q^2$ term in Eq.~\ref{eq:rate} to dominate at low $q^2$. 

The final term in Eq.~(\ref{eq:exp}) which is proportional to 
the nucleon density is the spin-averaged forward Compton 
scattering amplitude of virtual photons off a nucleon.  
Experimentally, data is only available for values of $q^2\leq0$, 
so while the photon rate which requires $q^2=0$ can be determined 
by use of the optical theorem the contribution to the dilepton 
rates must be determined by chiral constraints.  Broken chiral 
symmetry dictates uniquely the form of the strong interaction 
Lagrangian (at tree level) for spin-$\frac{1}{2}$ particles.  
Perturbative unitarity follows from an on-shell loop-expansion 
in $\frac{1}{f_\pi}$ that enforces current conservation and 
crossing symmetry.  To one-loop, the $\pi N$ contribution is 
parameter free.  The large contribution of the $\Delta$ to the 
Compton amplitude near threshold is readily taken into account by 
adding it as a unitarized tree term to the one-loop 
result~\cite{paper2,SZ99}.  The enhancement in the dimuon rates 
due to a non-vanishing baryon density can be seen in 
Fig.~\ref{fig:dimuon_rates_had} where the solid curve shows the 
total dimuon spectra with an enhancement as large as a factor of 
two in the invariant mass region of $2m_\mu \leq M \text{(GeV)} \leq 0.6$.

\section{Fireball}

As mentioned earlier, in order to compare the theoretical dilepton production 
rates with those seen in heavy-ion collisions it is necessary to integrate 
these rates over the space time evolution of the collision region.  We 
consider a region localized in space-time consisting of thermal hadronic 
matter acting as a source of particles.  Equilibrium of the collision 
region is strictly a local property with different temperatures and 
baryon densities possible in different space-time domains.  

A computational hydrodynamic code was already developed by one of us and 
it has been modified to the conditions of the SPS collider for Indium 
on Indium collisions.  In this paper we only briefly outline the physics 
behind this code and show the results of the In-In collisions which has 
not been modeled before.  For technical details regarding the hydrodynamic 
calculations the reader is referred to the prior works by one of 
us \cite{Teaney01}. 

\subsection{Hydrodynamics}

The hydrodynamic equations for a relativistic fluid consist of
the local conservation of energy and momentum, which can be written in 
compact form as $\partial_u T^{\mu\nu}=0$, as well as local 
charge conservation $\partial_\mu J^\mu_i=0$ where 
$T^{\mu\nu}=(\epsilon+p)U^\mu U^\nu-pg^{\mu\nu}$ is the 
energy-momentum tensor with $\epsilon$ the energy density, $p$ 
the pressure, $U^\mu=\gamma(1,{\bf v})$ is the proper velocity of 
the fluid, and $J^\mu_i$ is any conserved current ({\em e.g.} 
isospin, strangeness and baryon number in the case of strong interactions).

The same space-time evolution scenario as first proposed by 
Bjorken \cite{bjorken83} is assumed where the equation of motion 
can be described by the Bjorken proper time $\tau=\sqrt{t^2-z^2}$ 
and the spatial rapidity $y=\frac{1}{2}\ln\frac{t+z}{t-z}$.  One of 
the main results, following from the assumption of a central-plateau 
structure in the rapidity distribution is that of boost invariance, 
stating that the initial conditions and thus the subsequent evolution 
of the system are invariant with respect to a Lorentz boost.  Thus a 
solution at any value of $y$ can be found by boosting the solution at 
$y=0$ to a new frame moving with velocity $v=\tanh (y)$ in the negative 
z-direction.  

With the assumption of boost invariance the equations of motion are a 
function of the transverse coordinates and the proper time $\tau$ only.  
After integrating over the transverse plane of the collision region one 
finds that $(dS_{tot}/d\eta)$, $(dn_B/d\eta)$, and the net transverse 
momentum per unit rapidity are all conserved.

\subsection{Equation of State}

In order to solve the equations of motion as given by the vanishing of 
the divergence of the energy-momentum stress tensor one must have 
an Equation of State (EoS) relating the local values of the pressure, 
energy density, and baryon density ($n_B$).  The approach taken here 
is to consider an EoS with a variable latent heat in the $e/n_B$ plane 
where the following two derivatives hold along a path of constant $n_B/s$:

\begin{equation}
\label{eq:speed_sound}
\bigglb(\frac{dp}{de}\biggrb)_{n_B/s}\equiv c^2_s
\end{equation}

\begin{equation}
\label{eq:dsoverde}
\bigglb(\frac{ds}{de}\biggrb)_{n_B/s} = \frac{s}{p+e}
\end{equation}

If the speed of sound is defined everywhere along with the entropy of 
one arc in the $e,n_B$ plane the above derivatives can be integrated 
in order to determine $s(e,n_B)$.  From the entropy all other 
thermodynamic variables, such as T and $\mu_B$, can be found as needed. 

We consider a fireball that consists of three phases, a hadronic phase, 
a QGP phase, and a mixed phase.  The hadronic phase is taken to be made 
of ideal gas mixtures of the lowest SU(3) multiplets of mesons and 
baryons.  All intensive thermodynamic quantities including $p, e, s,$ 
and $n_B$ can be found as a sum of that quantity's contribution 
from each specie in the gas consisting of a simple Bose or 
Fermi distribution.  The hadronic phase is taken up to a temperature 
of $T_C \leq 170$ MeV and has a squared speed of sound of 
approximately $1/5c^2$.  For temperatures above $T_C$ only the 
squared speed of sound, $c^2_s$ is specified.  For the mixed phase 
it is taken almost at zero ($c_s^2 = 0.05c$). For the QGP phase the 
degrees of freedom are taken to be massless and the speed of sound is 
accordingly $c_s^2 = 1/3c$. The extension of this analysis to the
sQGP is beyond the scope of this work.

\subsection{Initial Conditions}

The initial conditions of the fireball consist of setting the 
entropy and baryon density proportional to the number of 
participating nucleons in the transverse plane at some 
initial proper time $\tau_0 = 1$ fm/c.  Since both the 
entropy and baryon number per unit rapidity are conserved 
the final yields of pions and nucleons are proportional to 
the number of participants.  The number of participants 
were calculated by use of a Glauber model and the initial 
entropy and baryon densities were fixed by two constants 
$C_s$ and $C_{n_B}$, which respectively are the entropy 
and net baryon number produced per unit spatial rapidity 
per participant.  These constants were fixed to the 
conditions at the CERN SPS collider in order to fit the 
total yield of charged particles and the net yield of 
protons.  Table~\ref{tab:param} summarizes the input 
parameters used in the hydrodynamic calculations. In order to address the centrality of the collision the 
impact parameter was chosen in order to reproduce the 
number of participants as reported in \cite{Usai}. 

\begin{table}
\begin{tabular}{l|c}
\hline
Parameter & Value \\
\hline
$c^2_{mixed}$	& 0.05c	\\
$c^2_{QGP}$	& 0.33c	\\
T$_C$	& 170 MeV\\
T$_{f.o.}$	&	130 MeV\\
$\tau_0$	&	1.0 fm/c\\
$n_B/s$	& 0.0238\\
$C_s$	&	8.06\\
$C_{n_B}$	&	0.191\\
\hline
\end{tabular}
\caption{\label{tab:param} Parameters used in the hydrodynamic 
simulation of In-In collisions.}
\end{table}

\subsection{Results for In-In Collisions at CERN SPS}

\begin{figure} 
\includegraphics[scale=1.2]{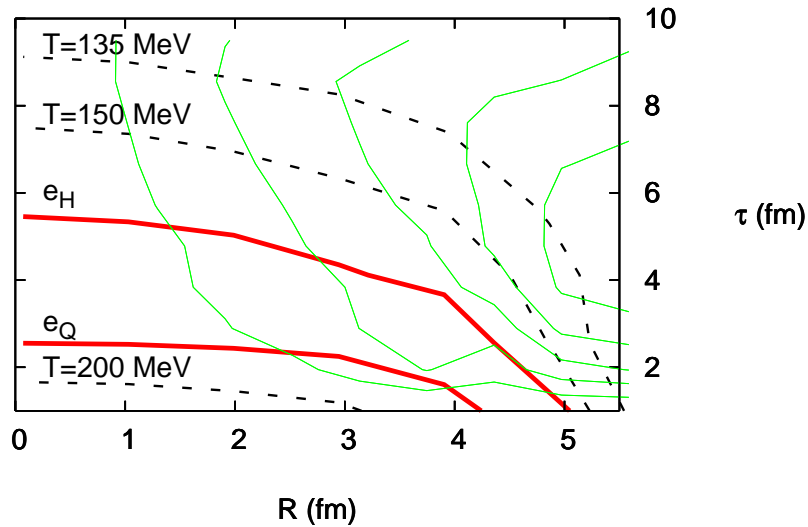}
\caption{The hydrodynamic solution for semi-central In-In collisions 
at the SPS collider.  The thin lines show contours of constant 
transverse fluid rapidity $(v_\perp=tanh(y_\perp))$ with 
values of 0.1,.02,..,0.5.  The dashed lines show contours of 
constant temperature with values of (working radially outward) 
T=200 MeV, T=150 MeV and T=135 Mev.  The $e_Q=1.70$ GeV/fm$^3$ 
and $e_H=0.50$ GeV/fm$^3$ contours represent the phase changes 
from QGP to mixed and from mixed to hadronic matter respectively.}
\label{fig:RvsTauSC}
\end{figure}

\begin{figure} 
\includegraphics[scale=.5]{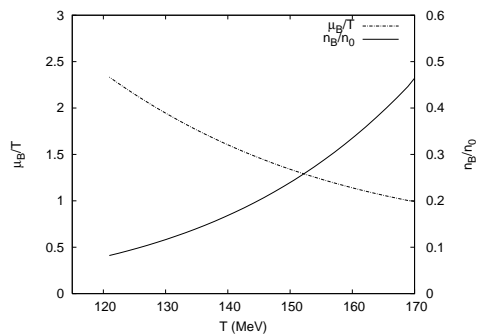}
\caption{Dependence of the baryon chemical potential, 
$\mu_B$, on the temperature for the hadronic phase of the fireball.}
\label{fig:muBvsT}
\end{figure}

The hydrodynamic result for In-In Semi-Central collisions is shown 
in Fig.~\ref{fig:RvsTauSC}.    The two thick lines labeled $e_Q$ and 
$e_H$ represent contours of constant energy density showing the 
transition from the plasma phase to the mixed plasma and hadronic 
phase and the transition from the mixed to the purely hadronic 
phase respectively.  It can be seen that the QGP phase takes up 
a much smaller space-time volume then the hadronic phase, however 
the rates still appear in the spectrum as the high temperatures 
in this region enhance the rates by an order of magnitude.  
The effect of nucleons depends on the baryon chemical potential 
in the fireball.  This is plotted as a function of temperature 
for the pure hadronic phase in Fig.~\ref{fig:muBvsT}. 

\section{Results and Discussion}

\begin{figure}[hbtp]\label{fig:result}
  \vspace{9pt}

  \centerline{\hbox{ \hspace{0.0in} 
\includegraphics[scale=.35]{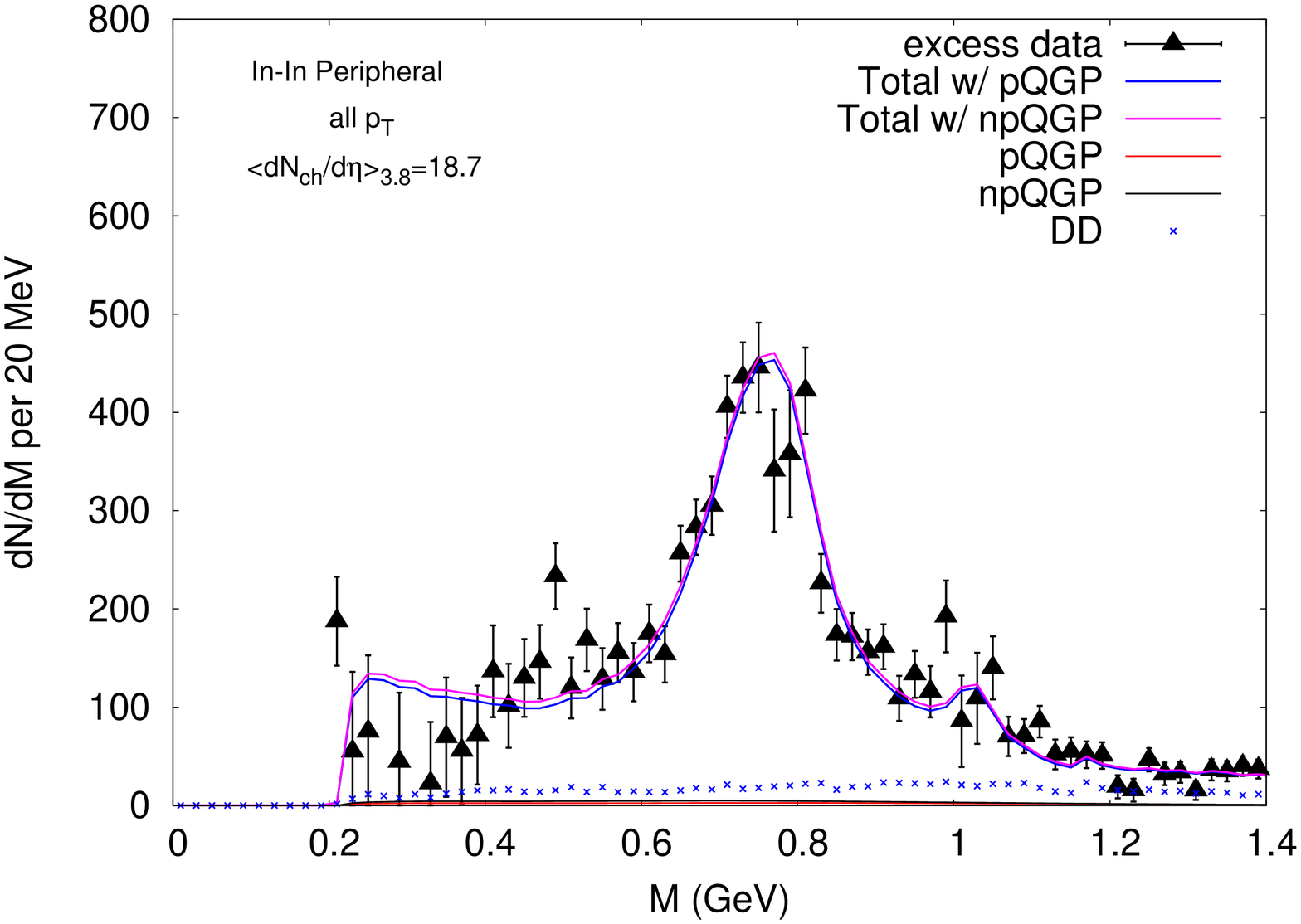}
    \hspace{0.1in}
\includegraphics[scale=.35]{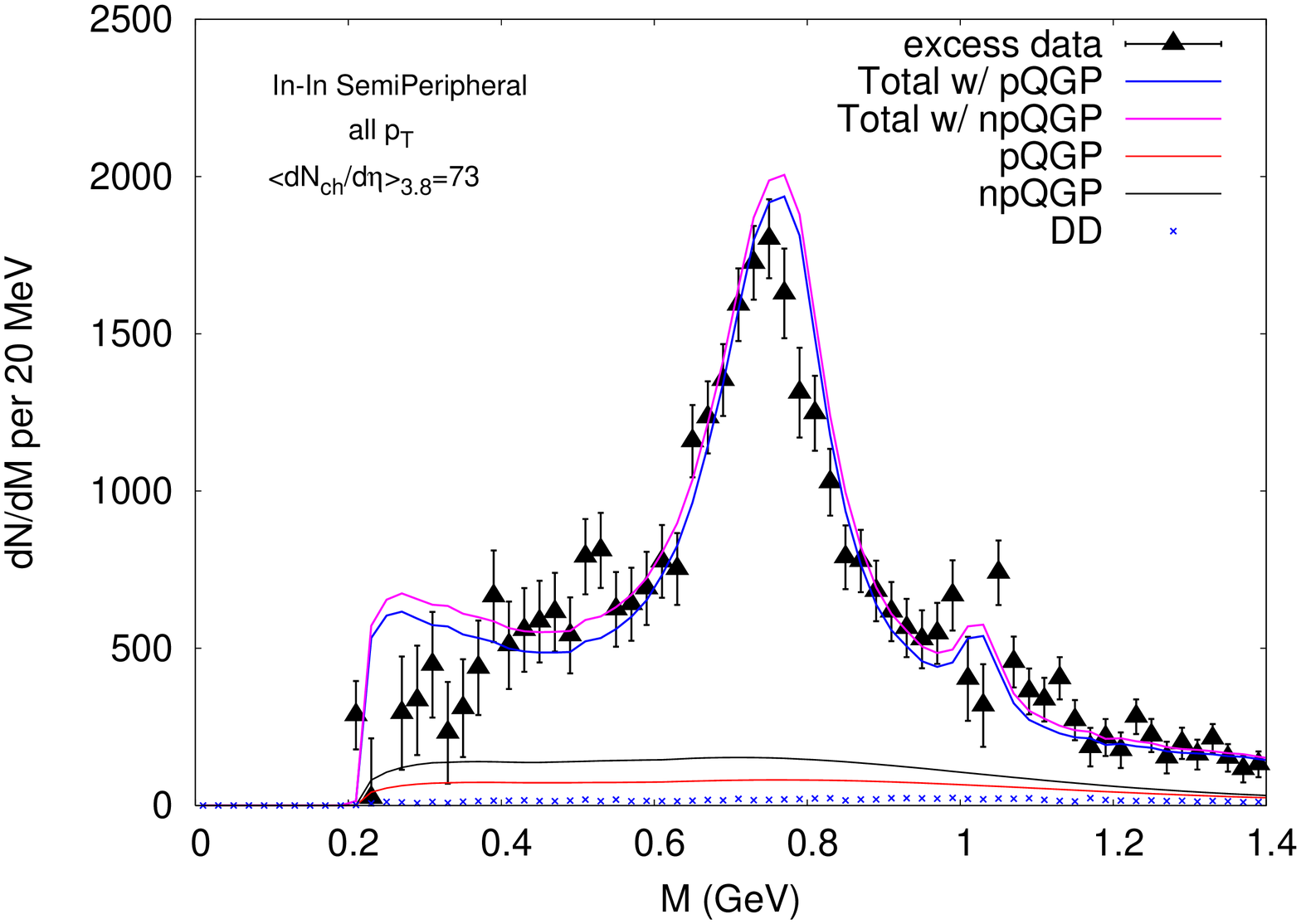}
    }
  }

  \vspace{9pt}

  \centerline{\hbox{ \hspace{0.0in}
\includegraphics[scale=.35]{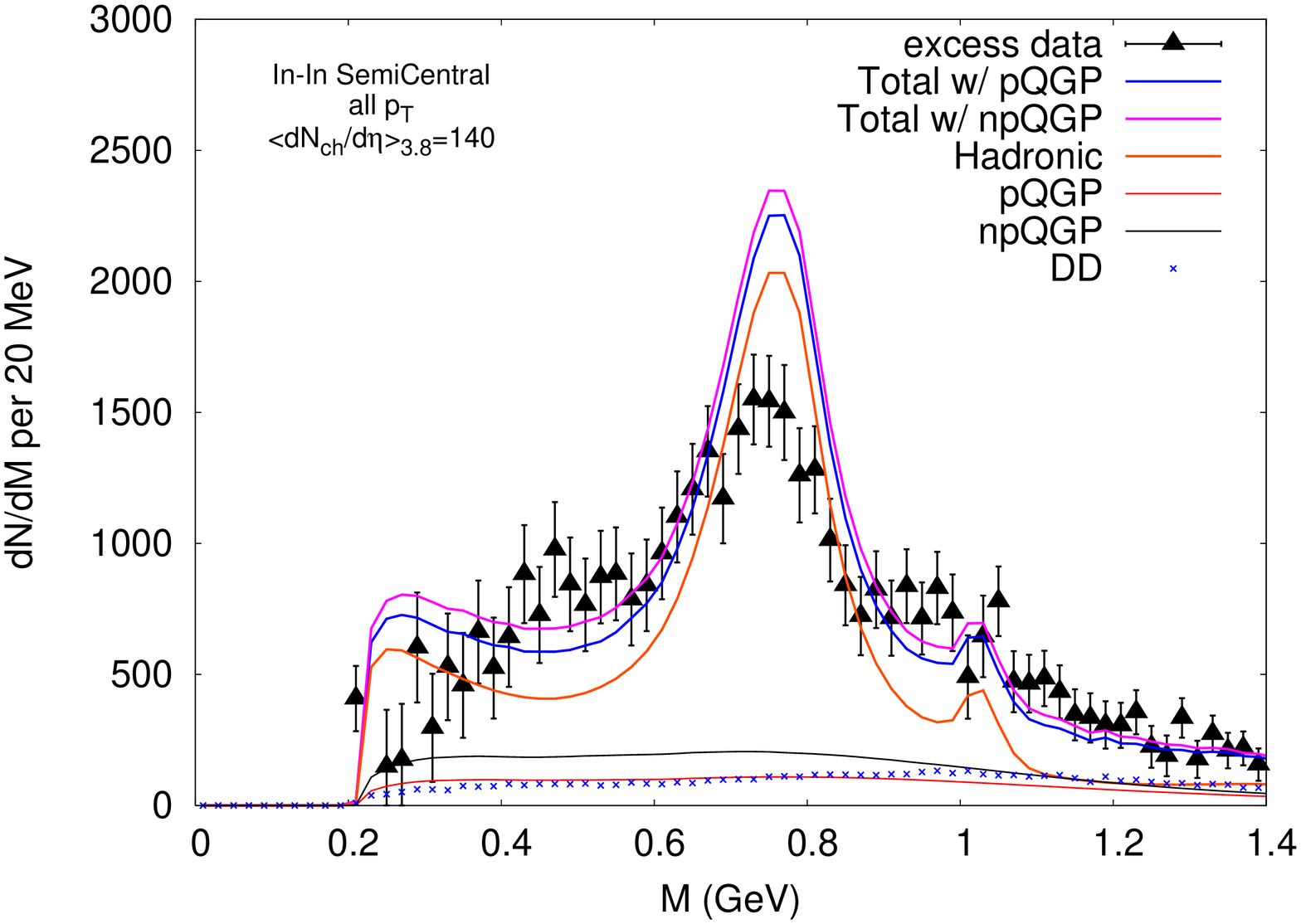}
    \hspace{0.1in}
\includegraphics[scale=.35]{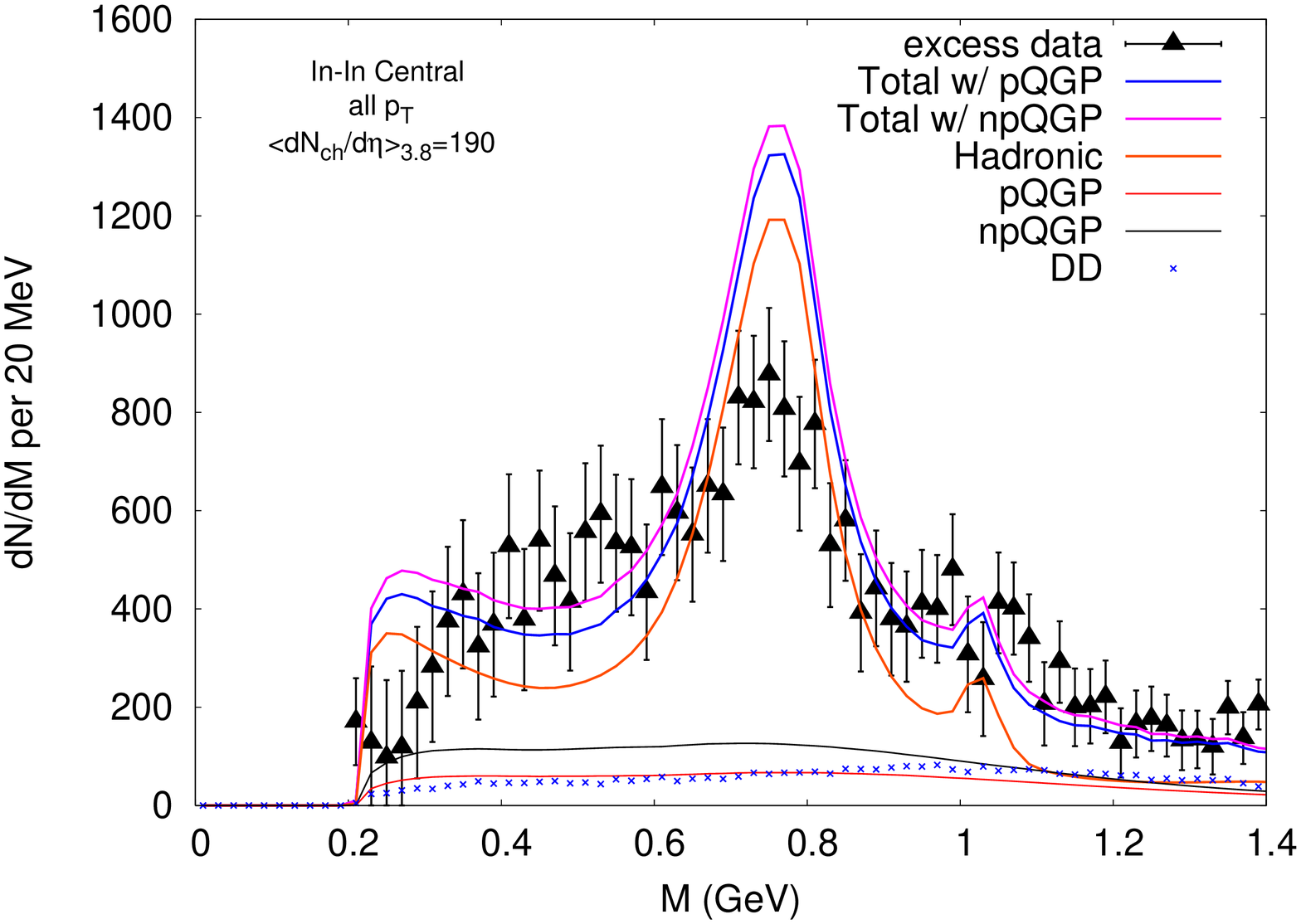}
    }
  }

  \vspace{9pt}

  \caption{ NA60's excess dimuon data compared to our thermal yields which include contributions from either the perturbative or non-perturbative QGP phase, the hadronic phase and the D\={D} contribution.  Shown for all four centrality windows.}
  \label{result}

\end{figure}

Our final dimuon yields for the four centrality windows is shown in 
Fig.~\ref{fig:result} where it is compared to the excess data measured by the 
NA60 collaboration.  In all four figures we show the total dimuon yield, which includes contributions from the hadronic phase, either the perturbative or non-perturbative plasma phase, as well as the D\={D} contribution as provided by the NA60 collaboration.  For all cases we also show separately the perturbative and non-perturbative plasma contributions to the overall yield.  It can be seen to be almost negligible in the peripheral data.  For the central data, where there is a larger plasma contribution we also show curves showing separately the hadronic contribution.
 
Even though it can 
be seen that the theoretical rates are able to describe most 
of the features of the spectrum, a number of things should be 
noted before a direct comparison is made.  The rates below M=0.4 GeV 
should not be taken literally since they are obtained by 
saturating the total measured yield in that region 
by $\eta$ Dalitz decays only, thereby lowering the 
excess close to the dimuon threshold.  Actually, by reducing 
the $\eta$ yield by 10\% the data has much better agreement with 
the theory for M$\leq0.4$ GeV.  The charm decay data was analyzed 
and provided by NA60, and since the contribution from charm 
decays is not subtracted from the excess data it must added to 
our rates for comparison with experiment.  The excess spectra 
which is shown in the figure is created from subtracting the 
cocktail (omitting the $\rho(770)$) from the total observed data.  
This would erase any $\omega$ or $\phi$ peak at the vacuum positions.  
Since our hadronic rates don't modify either the position or 
width of the $\omega$ or $\phi$ it can be very difficult to 
distinguish any residual $\omega$ or $\phi$'s from the cocktails'.  

It can be seen right away that the dimuon yields are reproduced in the peripheral centrality windows.  This is expected as the matter is dilute 
and any medium modification to the spectral densities will be accounted for in the virial expansion (Eq.~\ref{eq:exp}).   In the central bins it can be seen that the shape of the spectrum changes as one goes to more central collisions.  Even though the general shape of the spectrum is reproduced by our rates, our rates slightly over-predict the yield at the $\rho$ peak by about $\approx 50\%$ for semi-central and by $\approx 60\%$ for the central data.  Even though our rates agree fairly well with the remaining data away from the $\rho$ peak, there is still room for enhancement in the low mass region, $0.4 \leq \text{M (GeV)}  \leq 0.6$.

We should finally mention what happens when the non-perturbative 
QGP rate is used instead of the perturbative result.  Similar 
to the perturbative QGP results in Fig.~\ref{fig:result} 
the non-perturbative plasma rate is about a factor or two larger in the low mass region.  Even though this does help to explain
some of the excess in the low mass region, especially in the more central data, the space time volume of the plasma phase is too small
to have a large effect.

\section{Conclusions}

Using a parameterization of the results given by a hydrodynamic model 
of the collision region at the CERN SPS collider, the NA60 dimuon 
spectrum was reproduced using a pure thermal model assuming that 
there exists a sQGP phase above $T_C$ with an interacting hadronic 
phase persisting until freeze-out.  The dimuon spectrum from the sQGP 
phase originates primarily from q\={q} annihilation with non-perturbative 
effects due to non-vanishing gluon condensates.  After hadronization it 
is assumed that there remains a dilute hadronic gas in which the dimuon 
rates can be constrained entirely from broken chiral symmetry arguments and 
experimental data.  The combination of these two rate equations, 
after being folded over the space-time evolution of a fireball, 
are able to explain most of the excess dimuon data, especially in the more peripheral collisions where our 
assumptions about diluteness hold.  In the more central data, where the assumption of diluteness may 
breakdown, it is necessary to investigate how higher order terms in the virial expansion modify the spectrum.
\acknowledgments

We would like to thank Sanja Damjanovic for running our rates 
through the NA60 acceptance.  We are grateful to Gerry Brown, 
Axel Drees, Edward Shuryak, and Hans Specht for useful discussions.
This work was partially supported by the US-DOE grants 
DE-FG02-88ER40388 and DE-FG03-97ER4014.


\begin{thebibliography}{MM}

\bibitem{paper1} J. V. Steele, H. Yamagishi, 
and I. Zahed, Phys. Lett. B {\bf 384} (1997) 255.

\bibitem{paper2} J. V. Steele, H. Yamagishi, 
and I. Zahed, Phys. Rev. D {\bf 56} (1997) 5605; 
Nucl. Phys. A {\bf 638} (1998) 495c.

\bibitem{paper3} C. H. Lee, H. Yamagishi, 
and I. Zahed, Phys. Rev. C {\bf 58} (1998) 2899.

\bibitem{sanja05} S. Damjanovici, Parallel talk (exp) at QM 2005

\bibitem{HungShuryak96} C. M. Hung and E. V. Shuryak, 
Phys. Rev. C {\bf 56} (1997) 453.

\bibitem{CERES95} CERES Collaboration, 
G. Agakichev {\em et. al.,} Phys. Rev. Lett. {\bf 75}, 1272 (1995).

\bibitem{Rapp} R. Rapp, Nucl. Phys. A {\bf 725} 2003 254.

\bibitem{LiGale98} Guo-Qiang Li and C. Gale, 
Phys. Rev. C {\bf 58} (1998) 2914.

\bibitem{BrownRho}  G. E. Brown, M. Rho, Phys. Rev. Lett. {\bf 21}, 2720 (1991).

\bibitem{SZ99} J.V. Steele and I. Zahed, hep-ph/9901385 v2.

\bibitem{Hees06} Hendrik van Hees and Ralf Rapp, hep-ph/0603084.

\bibitem{Renk06} Thorsten Renk and J\"{o}rg Ruppert, hep-ph/0603110.

\bibitem{Huang95} Z. Huang, Phys. Lett. B {\bf 361}, 131 (1995).

\bibitem{McLerran} L. D. McLerran and 
T. Toimela, Phys. Rev. D {\bf 31}, 545 (1985).

\bibitem{Weldon} H. A. Weldon, Phys. Rev. D {\bf 42}, 2384 (1990).

\bibitem{PC} S. Damjanovic, private communication (2006).

\bibitem{Letessier18}  J. Letessier and J. Rafelski, 
{\em Hadrons and Quark-Gluon Plasma} (Cambridge University Press, 2002).

\bibitem{rapidity} NA60 Collaboration, M. Floris {\em et al.}, J.Phys.Conf.Ser.5:55-63,2005

\bibitem{bellac} M. Le Bellac, 
{\em Thermal Field Theory} (Cambridge University Press, 1996).

\bibitem{cleymans} J. Cleymans, 
J. Fingberg, and K. Redlich. Phys. Rev. D {\bf 35}, 7 (1987).

\bibitem{LWZH98} C.-H. Lee, J. Wirstam, 
I. Zahed, and T. H. Hansson, Phys. Lett. B {\bf 448}, 168 (1999).

\bibitem{CRF} H. Yamagishi and 
I. Zahed, Ann. Phys. (NY), {\bf 247}, 292 (1996).

\bibitem{bjorken83}  J. D. Bjorken, Phys. Rev. D {\bf 27}, 140 (1983).

\bibitem{Teaney01} D. Teaney, J. Lauret, and E.V. Shuryak, nucl-th/0110037.

\bibitem{Usai} NA60 Collaboration, G. Usai {\em et al.}, Eur. Phys. J. C (2005).

\end{thebibliography}
\end{document}